\documentclass[preprint,groupedaddress,showkeys]{revtex4}
\usepackage{epsf,graphicx}
\begin{document}

\title{The effect of Dirac phase on acoustic vortex in media with screw dislocation}

\author{Reza Torabi}
\thanks{Corresponding Author.\\ Email: rezatorabi@aut.ac.ir, Tel:+989125581265, Fax:+988626227814}
%\email{rezatorabi@aut.ac.ir, Tel:+989125581265, Fax:+988626227814}

\author{Zahra Rezaei}
\affiliation{Department of Physics, University of Tafresh, P.O.Box:
39518-79611, Tafresh, Iran}

\begin{abstract}

We study acoustic vortex in media with screw dislocation using the
Katanaev-Volovich theory of defects. It is shown that the screw
dislocation affects the beam's orbital angular momentum and changes
the acoustic vortex strength. This change is a manifestation of
topological Dirac phase and is robust against fluctuations in the
system.
\end{abstract}

 \maketitle

\maketitle

\section{Introduction}

The linearized equations of elasticity are analogous with Maxwell
equations \cite{Auld}. This analogy enables us to suggest new
phenomena for elastic waves by knowing their optical counterparts.
In particular, vortex phenomena in optics can be mapped onto
acoustic vortex.

An optical vortex beam focuses on rings rather than points and has
helical wavefront, Fig. 1, \cite{Curtis}.
\begin{figure}[ht]
\begin{center}
\includegraphics[scale=0.3]{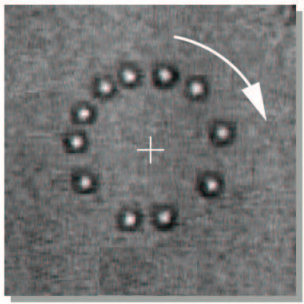}
\caption{The circular trajectory of a vortex beam \cite{Curtis}.}
\end{center}
\end{figure}
The difference between this kind of beam and a plane wave is just an
overall phase factor, $e^{il\varphi}$. The angle $\varphi$ is the
polar angle in cylindrical coordinates for a beam with axis parallel
to $z$ and $l$ is the optical vortex strength or the angular
momentum that is carried by the helical beam \cite{Bliokh,Mashhadi}.
When the helical beam interacts with a microscopic particle, the
orbital angular momentum can be transferred to the particle and make
it spin around the beam axis. A wide range of applications have been
recently found for this orbital angular momentum transfer. For
example we can mention, particle trapping \cite{Simpson} in optical
tweezers to manipulate micrometer-sized particles \cite{Curtis2} and
remote control of particles \cite{Shvedov}. Optical vortex also has
application in information encoding \cite{Marrucci}.

Acoustic vortex as the classical counterpart of optical vortex has
been studied \cite{Thomas,Marchiano,Thomas2,Zhang} and generated
\cite{Hefner, Gspan,Skeldon}, recently. Since acoustic vortices can
transfer orbital angular momentum to particles \cite{Skeldon,Volke},
like their optical counterparts, they can be applied to particle
trapping in acoustic tweezers and remote controlling, too. In
addition acoustic vortices, potentially, can be used in sonar
experiments \cite{Hefner}. Although similar properties to optical
vortex is expected for acoustic vortex, there are limited studies in
this area.

In this letter, acoustic vortex in media with screw dislocation is
studied using the Katanaev-Volovich theory of defects. The
motivation for studying defects is that they usually exist in
crystalline solids and have strong effect on their physical
properties
\cite{Mesaros,Teichler,Osipov,Vozmediano,Yazyev,Bausch1,Bausch2,Furtado1,Furtado2}.
In the presence of defects, we are confronting with complicated
boundary conditions. This difficulty persuades physicists to
introduce new approaches such as Katanaev-Volovich theory of defects
in solids
\cite{Furtado1,Furtado2,Furtado3,Katanaev,Katanaev2,Ali,Kleinert}.
Katanaev-Volovich theory is a geometrical approach based on the
isomorphism existing between the theory of defects in solids and
three-dimensional gravity. In this formalism, elastic deformation
which is introduced in the medium by defects is replaced by a
non-Euclidean metric. According to this theory, at distances much
larger than the lattice spacing where the continuum limit is valid,
the solid can be described by a Riemann-Cartan manifold.
Dislocations and disclinations of the medium are respectively
associated with torsion and curvature of the manifold. We will show
that the screw dislocation changes the acoustic vortex strength.
This change is due to Dirac phase and is robust against
fluctuations. Dirac phase belongs in the category of non-integrable
phase factors that appear in many different areas of Physics
\cite{Shapere,Mehrafarin,Torabi}. Dirac showed that when a particle
transports in an external electromagnetic field, its wave function
acquires a phase term in addition to usual dynamic phase factor
\cite{Dirac,Bliokh1}.

The letter is organized as follow. In section II, we review the
screw dislocation in Katanaev-Volovich formalism. Section III is
devoted to the Dirac phase of acoustic waves in media with screw
dislocation. The effect of noise on Dirac phase is discussed in
section IV. Finally the conclusion is presented in section V.

\section{Screw dislocation in Katanaev-Volovich formalism}

Consider a point in an undeformed medium with coordinates $x^i$ with
Euclidean metric $\delta_{ij}$. If the deformation due to the defect
is described by a displacement vector $U^i(x)$, the point will have
coordinates $y^i=x^i+U^i(x)$. According to the Katanaev-Volovich
approach the effect of the elastic deformation $U^i$ is considered
by introducing metric $g_{ij}$ which can be expressed in terms of
the initial metric $\delta_{ij}$ as \cite{Katanaev,Katanaev2,Ali}
\[
g_{ij}:=\frac{\partial x^k}{\partial y^i}\frac{\partial
x^l}{\partial y^j}\delta_{kl},\;\;\;\; i,j=1,2,3.
\]
One of the defects, which we are interested in, is a screw
dislocation. In a screw dislocation the Burgers vector is parallel
to the dislocation line. This kind of defect, which corresponds to a
singular torsion along the defect line, is described by the
following metric \cite{Furtado3,Tod}
\[
ds^{2} =g_{ij}dx^{i}dx^{j}= \left( dz + \beta d \phi \right)^{2} + d
\rho^{2} + \rho^{2} d \phi^{2},
\]
where the parameter $\beta$ is related to the Burgers vector,
$\bf{b}$, by $\beta = \frac{b}{2 \pi}$ and the screw dislocation
line is oriented along the z-axis of the cylindrical coordinates
$(\rho, \varphi, z)$. To derive the above metric, we have used the
displacement vector ${\bf U}=(0,0,\beta\phi)$ associated with a
screw dislocation in cylindrical coordinates. The metric tensor
$g_{ij}$ is
\begin{equation}
\label{1}
g_{ij}=\left( \begin{array}{ccc} 1 & 0 & 0 \\
0 & \beta^2+\rho^2 & \beta \\
0 & \beta & 1 \end{array} \right)
\end{equation}
and carries no curvature.

The torsion two-form associated with this defect is defined as
$T^a=T_{ij}^a dx^i \wedge dx^j$, ($a\equiv\{\rho,\phi,z\}$), that
the only non-vanishing component is given by \cite{Furtado2}
\[
T^z=2\pi \beta \delta ^2(\rho )d\rho \wedge d\phi,
\]
where $\delta ^2(\rho )$ is the two-dimensional delta function in
flat space and reveals the singularity in torsion. Also the torsion
in tensor notation can be written as
\begin{equation}
\label{2} T_{ij}^a =\partial _i e_j ^a -\partial _j e_i ^a,
\end{equation}
where $e_i^a$ are triad components. Comparison of equation (2) with
the field strength $F_{ij} =\partial _i A_j -\partial _j A_i $ in
the electromagnetism and the singular value of the torsion field,
indicates a similarity between this case and the Aharanov-Bohm
effect \cite{Aharonov}, where the Burgers vector plays the role of
the magnetic flux.

\section{Dirac phase of acoustic waves}

The dynamic of displacement vector field ${\bf U}({\bf x},t)$ in an
elastic medium without defect is governed by (see e.g.
\cite{Landau})
\begin{equation}
\label{3}
\partial _t^2 U^{i}=\frac{\mu }{\rho }\nabla ^2 U^{i}+\frac{(\lambda
+\mu )}{\rho }{\partial}^{i}{\partial}_{j}U^{j},
\end{equation}
where $\lambda $ and $\mu $ are the Lame coefficients and $\rho$ is
the density of the medium. According to the Katanaev-Volovich
approach, media with defects can be treated by nontrivial metric,
$g_{ij}$. Therefore, the covariant generalization of equation (3)
gives the displacement vector dynamics in media with defects
\begin{equation}
\label{4}
\partial _t^2 U^{i}=\frac{\mu }{\rho }\tilde{\nabla} ^2 U^{i}+\frac{(\lambda
+\mu )}{\rho }\tilde{\nabla}^{i}\tilde{\nabla}_{j}U^{j}.
\end{equation}
The displacement vector can be decomposed covariantly into
transversal, $U^{Ti}$, and longitudinal, $U^{Li}$, parts
\cite{Katanaev2},
\[
U^{i}=U^{Ti}+U^{Li},
\]
which satisfy the following relations
\[
\tilde{\nabla}_{i}U^{Ti}=0,
\]
\[
\tilde{\nabla}_{i}U^{L}_{j}-\tilde{\nabla}_{j}U^{L}_{i}=0.
\]
As far as we know from vector analysis, it is always possible to
express a vector as the sum of the curl of a vector and the gradient
of a scalar. So equation (4) decomposes into two independent
equations for transverse and longitudinal parts of the displacement
vector field,
\begin{equation}
\label{5} \frac{1}{v^2_T}\partial _t^2 U^{Ti}-\tilde{\nabla} ^2
U^{Ti}=0,\;\;\;\;\;\;\frac{1}{v^2_L}\partial _t^2
U^{Li}-\tilde{\nabla} ^2 U^{Li}=0,
\end{equation}
where
\[
v^2_T=\frac{\mu }{\rho },\;\;\;\;\;v^2_L=\frac{(\lambda +2\mu
)}{\rho},
\]
are the speeds of transverse and longitudinal parts in the medium.
So by decomposition of (4) every longitudinal or transverse
component of the displacement vector, $\bf U$, satisfies a separate
scalar wave equation in the curved space. $\tilde{\nabla}^{2}$ in
(5) is the Laplace-Beltrami operator which is given by
$\tilde{\nabla}^2=\frac{1}{\sqrt{g}}\partial_i(g^{ij}\sqrt{g}\partial_j)$,
where $g$ is the determinant of the metric tensor $g_{ij}$ and
$g^{ij}=(g_{ij})^{-1}$ is its inverse. This decomposition enables us
to study each mode separately. Here we are interested in the
longitudinal mode but similar results can be deduced for the
transverse mode, too.

Using the metric tensor (1) for screw dislocation, the longitudinal
part of the elastic wave in (5) takes the form
\[
\bigg{\{}\frac{1}{\rho}\partial_\rho(\rho\partial_\rho)+\frac{1}{\rho^2}(\partial_\varphi-\beta\partial_z)^2+\partial_z^2\bigg{\}}
U^{Li}(\rho, \varphi, z,t)=\frac{1}{v^2_L}\partial _t^2 U^{Li}(\rho,
\varphi, z,t).
\]
Considering a monochromatic paraxial wave as
\[
U^{Li}(\rho,\varphi,z,t)=e^{-i\omega t}e^{ikz}u^{Li}(\rho,\varphi),
\]
yields to the following equation for longitudinal elastic wave
\begin{equation}
\label{6}
\bigg{\{}\frac{1}{\rho}\partial_\rho(\rho\partial_\rho)+\frac{1}{\rho^2}(\partial_\varphi-i\beta
k)^2-k^2\bigg{\}} u^{Li}(\rho,
\varphi)=-\frac{\omega^2}{v^2_L}u^{Li}(\rho, \varphi).
\end{equation}
Equation (6) implies that
$\partial_\varphi\rightarrow\partial_\varphi\ -ik\beta$ with respect
to the defect free case ($\beta=0$) in which the Laplacian operator
is given in a flat space. In the other words, the $z$-component of
angular momentum has changed according to $L_z\rightarrow\ L_z -
k\beta$. The angular momentum of the acoustic wave along it's axis
is modified by the presence of the defect that is due to the torque
exerted by the strain field of the dislocation. Introducing a
momentum operator as ${\bf P}=-i\nabla$ converts (6) into a
time-independent Schr\"{o}dinger-like equation with a gauge
potential
\begin{equation}
\label{7} ({\bf P}-{\bf A})^2 u^{Li}(\rho,
\varphi)=\frac{\omega^2}{v^2_L} u^{Li}(\rho, \varphi),
\end{equation}
where the corresponding vector gauge potential is
\begin{equation}
\label{8} \mathbf{A}=\frac{k\beta}{\rho} \hat{\mathbf{e}}_\varphi.
\end{equation}
Since this gauge is curl free, $\nabla \times {\bf A}=0$, the
perfect analogy is seen between acoustic waves in media with screw
dislocation and the Aharanov-Bohm effect. Note that, the torsion
field is invariant under gauge transformations of the potential,
\[
\mathbf{A}\rightarrow \mathbf{A}+\nabla\Lambda.
\]
According to this correspondence, Dirac phase factor method
\cite{Dirac,Sakurai} can be used here (See the appendix). Thus, the
solution of the wave equation (7) has the following property
\begin{equation}
\label{9} u^{Li}(\rho,\varphi)=\exp\bigg{\{}i\int_C \mathbf{A}\cdot
d\mathbf{r}\bigg{\}}u_0^{Li}(\rho,\varphi),
\end{equation}
where $u_0^{Ti}(\rho,\varphi)$ is the solution of the defect free
case and $C$ is the beam trajectory (See Fig. 1). Substituting (8)
into (9) yields
\begin{equation}
\label{10} u^{Li}(\rho,\varphi)=e^{i\int_{0}^{\varphi}k\beta
d\varphi}u_0^{Li}(\rho,\varphi).
\end{equation}
This means that $u^{Li}(\rho,\varphi)$ differs from
$u_0^{Li}(\rho,\varphi)$ just in a phase factor $e^{i\gamma}$ that
\[
\gamma=\int_{0}^{\varphi}k\beta d\varphi,
\]
is called Dirac phase. In other words, the coupling of torsion with
angular momentum leads to an additional phase factor in the solution
when the screw dislocation is present.

Hitherto we found that the difference between the solutions to the
acoustic wave equation in the presence and in the absence of defects
is just manifested in a phase factor, (10). So we only need to find
the solution for the defect free case. In this case, $\beta=0$, the
solution of the wave equation (6) can be easily found as
\begin{equation}
\label{11} u_0^{Li}(\rho,\varphi)=R(\rho)e^{il\varphi},
\end{equation}
where $R(\rho)$ is the radial solution of the Helmholtz equation and
$e^{il\varphi}$ represents acoustic vortex carrying the angular
momentum $l$, acoustic vortex strength, along the paraxial axis.
According to equations (10) and (11) the solution of the wave
equation in the presence of screw dislocation is shown as
\[
u^{Li}(\rho,\varphi)=R(\rho)e^{i(l+\beta k)\varphi}.
\]
Therefore, the screw dislocation results in the change of the
acoustic vortex strength from $l$ to $l+\beta k$. This change, due
to Dirac phase, is proportional to the magnitude of Burgers vector
or in other words the flux of torsion.

\section{The effect of noise on Dirac phase}

The presence of noise is an inevitable subject in physical systems,
such as the the ubiquitous thermal fluctuation. The effect of noise
on the Dirac phase can be treated similar to the problem of
electrons in media with screw dislocations \cite{Torabi1}. The same
procedure is used to show that considering a white noise leads to
the average zero for the Dirac phase. This kind of noise coincides
with uncorrelated nature of thermal noise. Indeed, the variance of
the Dirac phase diminishes with time as
\[
\langle\triangle\gamma^{2}(T)\rangle\propto\frac{1}{T}
\]
where $T$ is the period of the beam's rotation on its circular
trajectory (Fig. 1). Therefore, $\gamma$ coincides with its
noiseless value in the limit $T\rightarrow\infty$. As a result, in
spite of the dynamic phase, the Dirac phase of elastic waves is
robust against fluctuations in the system.

\section{Conclusion}

In this letter we studied the effect of the screw dislocation on an
acoustic wave using the Katanaev-Volovich theory of defects. This
theory which is a geometrical approach uses the isomorphism between
the theory of solids and three-dimensional gravity to suppress the
technical complications due to adding defects to the system. It was
shown that the screw dislocation changes the acoustic vortex
strength by coupling of torsion with orbital angular momentum. This
change is a manifestation of the topological Dirac phase and is
robust against fluctuations. For a white noise, coincides with the
nature of thermal noise, the effect of fluctuations on the Dirac
phase diminishes as $\frac{1}{T}$ where $T$ is the period of beam's
rotation.

\textbf{Appendix: Dirac phase factor method}

Consider two Schr\"{o}dinger-like wave equations with two different
gauge fields ${\bf A}_1$ , ${\bf A}_2={\bf A}_1+\nabla \Lambda$
\begin{equation}
\label{A1} ({\bf P}-{\bf A}_1)^2 u_1^{Li}(\rho,
\varphi)=\frac{\omega^2}{v^2_L} u_1^{Li}(\rho, \varphi),
\end{equation}
and
\begin{equation}
\label{A2} ({\bf P}-{\bf A}_1-\nabla \Lambda)^2 u_2^{Li}(\rho,
\varphi)=\frac{\omega^2}{v^2_L} u_2^{Li}(\rho, \varphi).
\end{equation}
These wave equations are equal due to the gauge invariance, so we
are going to find the relation between $u_1^{Li}$ and $u_2^{Li}$ to
make this equality possible. We claim that
\begin{equation}
\label{A3} u_2^{Li}=e^{i\Lambda({\bf x})}u_1^{Li},
\end{equation}
and put it in (13) which yields to
\[
({\bf P}-{\bf A}_1-\nabla \Lambda)^2 e^{i\Lambda({\bf
x})}u_1^{Li}(\rho, \varphi)=\frac{\omega^2}{v^2_L} e^{i\Lambda({\bf
x})}u_1^{Li}(\rho, \varphi).
\]
Rewriting the last equation as follow
\[
e^{-i\Lambda({\bf x})}({\bf P}-{\bf A}_1-\nabla
\Lambda)e^{i\Lambda({\bf x})}e^{-i\Lambda({\bf x})}({\bf P}-{\bf
A}_1-\nabla \Lambda) e^{i\Lambda({\bf x})}u_1^{Li}(\rho,
\varphi)=\frac{\omega^2}{v^2_L} u_1^{Li}(\rho, \varphi).
\]
and comparing with (12) leads to
\[
e^{-i\Lambda({\bf x})}({\bf P}-{\bf A}_1-\nabla \Lambda({\bf
x}))e^{i\Lambda({\bf x})}={\bf P}-{\bf A}_1.
\]
Putting ${\bf P}=-i\nabla$ will complete the proof. If we consider
(12) for a defect free case, ${\bf A}_1=0$, and (13) for a case with
defect, ${\bf A}_2=\bf A=\nabla \Lambda({\bf x})$, then according to
(14)
\[
u_2^{Li}=\exp\bigg{\{}i\int_C \mathbf{A}\cdot
d\mathbf{r}\bigg{\}}u_1^{Li}.
\]
Therefore the two solutions are related by a phase factor which is a
function of the gauge field.

\end{document}